\documentclass{acmtrans2e} 
\RequirePackage{epsf} 
\RequirePackage{listing}

\begin{document}
	
\raggedbottom	

\title{Perpetual Adaptation of Software to Hardware: \\
An Extensible Architecture for Providing Code  \\ 
Optimization as a Central System Service \\} 

\author{THOMAS KISTLER and MICHAEL FRANZ \\
University of California, Irvine}

\begin{abstract}
Much of the software in everyday operation is not making optimal use of the hardware
on which it actually runs, but has been optimized for earlier and often outdated
processor versions. This is because users upgrade hardware independently from, and
often more frequently than, application software. Moreover, software vendors are
generally unable or unwilling to provide multiple versions of the same program that
differ only in the processor model specifically targeted during compilation.

The obvious solution to matching a piece of software with the actual capabilities
of the hardware on which it is about to be executed is to delay code generation until
load time. This is the earliest point at which the software can be fine-tuned to
specific hardware characteristics such as the latencies of individual instructions
and the sizes of the instruction and data caches. An even better match can be achieved
by replacing the already executing software at regular intervals by new versions
constructed on-the-fly using a background code re-optimizer. The code produced using
such dynamic re-optimizers is often of a higher quality than can be achieved using
static ``off-line'' compilation because live profiling data can be used to guide
optimization decisions and the software can hence adapt to changing usage patterns
and the late addition of dynamic link libraries.

Previous discussions of run-time code generation and optimization have focused on
functional aspects and on specific optimization techniques. This paper instead concentrates
on structural and conceptual aspects of such systems. Based on an actual implementation,
we present the architecture of a system that provides continuous application profiling,
continuous background re-optimization guided by the collected profiling information,
and continuous replacement of already running application software by re-optimized
versions of the same software.

A central trait of our architecture is extensibility. The dynamic optimizer at the
heart of our system has a component structure supporting incremental modification
in a plug-and-play manner. This is an essential facility for making system-level
code generation useful in practice. Without this capability, the ``outdated software''
problem would merely be shifted downward to the system level: system software manufacturers
would be just as disinclined to provide multiple run-time system versions as application
software manufacturers are unwilling to provide multiple versions of their products.
Among the conceptual issues discussed in the paper are the questions of when to trigger
re-optimizations, and which parts of the running software to re-optimize.
\end{abstract}

\category{D.3.4}{Programming Languages}{Processors}[Run-time Environments]
\category{D.3.4}{Programming Languages}{Processors}[Code Generation] 
\category{D.3.4}{Programming Languages}{Processors}[Compilers] 
\category{D.3.4}{Programming Languages}{Processors}[Optimization]

\terms{Dynamic Code Generation, Dynamic Re-Optimization, Extensibility}
\enlargethispage{\baselineskip}
\maketitle

\begin{bottomstuff}
Authors' address: Department of Information and Computer Science, University of California at Irvine, Irvine, 
CA 92697--3425. Part of this work is funded by the National Science Foundation under grant 
CCR--97014000.  A precursor to the chapter on the similarity of profiling data previously appeared as a refereed contribution 
in the \emph{ Proceedings of the Workshop on Profile and Feedback-Directed Optimization}, Paris, France, October 
1998.
\end{bottomstuff}

\section{Introduction} 
In the wake of dramatic improvements in processor speed, it is often overlooked that
much of the software in everyday operation is not making the best use of the hardware
on which it actually runs. The vast majority of computers are either running application
programs that have been optimized for earlier versions of the target architecture,
or, worse still, are emulating an entirely different architecture in order to support
legacy code. More recently, architecture emulation has also become widespread in
software portability schemes such as the Java Virtual Machine \cite{LY97}.

There are many reasons for the common mismatch between the processor for which a
piece of software was originally compiled and the actual host computer on which it
is eventually run. None of these reasons is fundamentally technical in nature. Partly,
it is the users' fault, who demand backward compatibility and are unwilling to give
up their existing software when upgrading their hardware. As a result of this, an
immense amount of legacy code is in use every day: 16-bit software on 32-bit processors,
emulated MC680x0 code on PowerPC Macintosh computers, and soon also IA32 code on
IA64 hardware.

But it is not only the users who are to blame: purely logistic constraints make it
unfeasible for software vendors to provide separate versions of every program for
every particular hardware implementation of a processor architecture. Just consider:
there are several major manufacturers of IA32-compatible CPUs, and each of these
has a product line spanning several processors---the total variability is far too
great to manage in a centralized fashion.

Furthermore, availability of production compilers is often lagging behind advances
in hardware, so that a software provider sometimes has no choice but to target an
outdated processor version. Take Intel's MMX technology as an example: The MMX instruction-set
was introduced in January 1997 to improve performance of media-rich applications.
Two years later, there still is no automatic support for MMX instructions in major
production compilers. In order to make use of these new instructions, software developers
need to laboriously rewrite their applications, inserting the appropriate calls manually.
Hence, it comes as no surprise that the impact of MMX has been negligible so far:
only a tiny fraction of programs exploits the special multimedia capabilities of
MMX processors---mainly specially handcrafted applications, written largely in assembly
language.

The unpleasant reality that non-technical circumstances can lead to actual performance
penalties is a direct consequence of the current practice of compiling applications
way ahead of their use. At compile time, each program is custom-tailored for a particular
set of architectural features, such as a particular instruction set with specific
latencies, a particular number of execution units, and particular sizes of the instruction
and data caches. The program is then constrained by these choices and cannot adapt
to hardware changes occurring only after compilation.

The obvious solution to overcoming these limitations is to delay the binding of architectural
information until it is certain that no further changes will occur. In most cases,
the earliest point at which a specific set of hardware characteristics becomes definite
for a certain piece of software is when the software is loaded into main memory immediately
prior to execution. Hence, the problem of matching the software to the hardware can
be solved in a straightforward manner simply by deferring native code generation
at least until load time. This approach has been validated in previous work and is
now common practice \cite{DS84,Fra94,Hol94,HU96,ACL+98}.  Besides enabling a closer match between software 
and hardware, load-time code generation also makes it possible to perform optimizations across the boundaries 
of independently-distributed software components and hence can reduce the performance penalty paid for 
modularization.

Interestingly enough, a still better match between the software and the hardware
can be achieved by re-evaluating the bindings at regular intervals instead of permanently
fixing them at load time. To this effect, a profiler constantly observes the running
program and determines where the most effort is spent. Using this information as
a guide, a new version of the already running software is then constructed on-the-fly
in the background, placing special emphasis on optimizing the critical regions of
the program. When the background re-optimization is complete, the new software is
``hot-swapped'' into the foreground and execution resumes using the new code image
rather than the old one. The latter can be discarded as soon as the last thread of
execution has been migrated away from it.

Because of its feedback loop, object code produced by dynamic re-optimization is
often of a higher quality than can be achieved using static ``off-line'' compilation.
Contrast this to load-time code generation, which in theory can be equally good as
static compilation, but which in practice is constrained by the time that is available
for code generation. In the presence of interactive users, this time is often severely
limited. Re-optimization doesn't have these constraints, because it happens strictly
in the background, while an alternate version of the application program is already
executing. Consequently, the speed of re-optimization is almost completely irrelevant;
even re-optimization cycles that last on the order of 10 minutes are still useful.

Dynamic re-optimization also presents an elegant solution to the problem of providing
\emph{dynamic loading} of user-level program extensions (dynamic link libraries) without having to pay a 
permanent performance penalty for this capability.  Many traditional optimizations cannot usually be applied 
in the context of dynamic loading.  For example, a method invocation in an object-oriented language can be 
statically bound at the call site if class hierarchy analysis indicates that the method is not overridden in 
any subtype \cite{Fer95,DGC95}.  However, such an optimization would normally be illegal if it were possible to load a 
new class at run-time that contained an overriding method.  In a system offering dynamic re-optimization, the 
more efficient statically bound call can be generated---as long as the system can guarantee that this 
optimization will be undone if overriding should occur at some later point.  In a more general sense, the 
ability to re-optimize the whole system even after it has already started running makes it possible to adapt 
to the presence of dynamic link libraries as they are loaded, and optimize each one in the context of all the 
others, not just the ones loaded previously.

In the re-optimization model, code generation and code optimization become central
services of the run-time system \cite{Fra97}.  The advantages are apparent: not only does this solution provide 
the highest possible code quality, but it also enables every application program to take full advantage of the 
actual hardware---provided that the run-time system itself has been targeted toward exactly this hardware and 
contains the appropriate dynamic optimizer.  Exactly there's the rub: how do we make sure that we haven't 
merely shifted the original problem of matching hardware and software ``one level downward'', into the 
run-time system?

It is this question that this paper presents an answer to. While previous work has
focused mainly on the functionality and benefits of dynamic code optimization---that
is, which types of optimizations and which profiling services should be pro\-vid\-ed---the
work presented here concentrates on the architectural aspects of such a system: In
what manner should dynamic profiling and dynamic optimization services be provided,
so that the creator of the run-time system need not simultaneously support separate
versions for every conceivable variation of the hardware architecture? Clearly, the
underlying run-time system cannot be structured as a monolithic kernel.

We have implemented a model run-time system providing dynamic profiling, dynamic
optimization, and dynamic replacement of live code and data and describe its architecture
below. Our system particularly distinguishes itself by its extensibility. Both the
dynamic profiler and the dynamic optimizer at the heart of our system have a component
structure supporting incremental modification in a plug-and-play manner. Users migrating
to a new set of hardware features then merely need the appropriate plug-in components
that match the new target architecture, rather than a whole new run-time system.
These plug-in components are only loosely coupled to the rest of the run-time system
and communicate with it via a message bus.

The central idea behind our architecture is that hardware designers know a lot about
their specific product, but relatively little about run-time systems in general.
A plug-in component for a run-time optimizer is akin to a device driver, except that
it enables an application program to utilize the main computing engine more effectively.
Just as operating systems today are shipped with a large number of device drivers
for every conceivable piece of hardware that an end-user might want to install, an
operating system incorporating an extensible code optimizer at its core would rely
on a set of plug-in optimization components supplied by the manufacturers of the
various processors. Hence, instead of today's centralized approach, in which software
suppliers need to keep track of, and maintain appropriate compilers for, all the
architectures for which they want to provide optimized code, our solution shifts this
responsibility to the hardware providers, completely eliminating the problems
of hardware variability mentioned above.

Indeed, it is entirely within the scope of our architecture to support even coprocessors
such as graphics accelerators, digital signal processors, and multipurpose FPGA processor
boards. In this case, the appropriate optimization plug-in would map calls of the
relevant standardized API functions directly into the specific hardware instructions
of the coprocessor and emit them as in-line code.

\enlargethispage{\baselineskip}
The remainder of this paper is organized as follows: Section 2 gives a brief overview of our component-based 
architecture in which individual parts, such as profilers and optimizers, are highly customizable and can be 
exchanged, removed, and added arbitrarily.  Sections 3 through 6 discuss individual components of the system, 
namely the profiling and optimization frameworks, and the code replacer.  Section 7 addresses open problems.  
Section 8 discusses related work and Section 9 concludes the paper.

\section{Architectural Overview} 
As illustrated in Figure 1, our dynamic code generation and optimization system is
composed of five main constituents: a manager, a code generating loader, a profiler,
an optimizer, and a replacer. This assembly of sub-systems is in turn part of a larger
run-time system that provides many additional services. These further capabilities
of the run-time system will not be discussed in this paper, but in order to support
the architecture described here, they must necessarily include dynamic loading of
software modules, run-time type-tagging of dynamically allocated objects, and garbage
collection.

\begin{figure}[htb]
\epsfxsize=4.5 in
\centerline{\epsffile{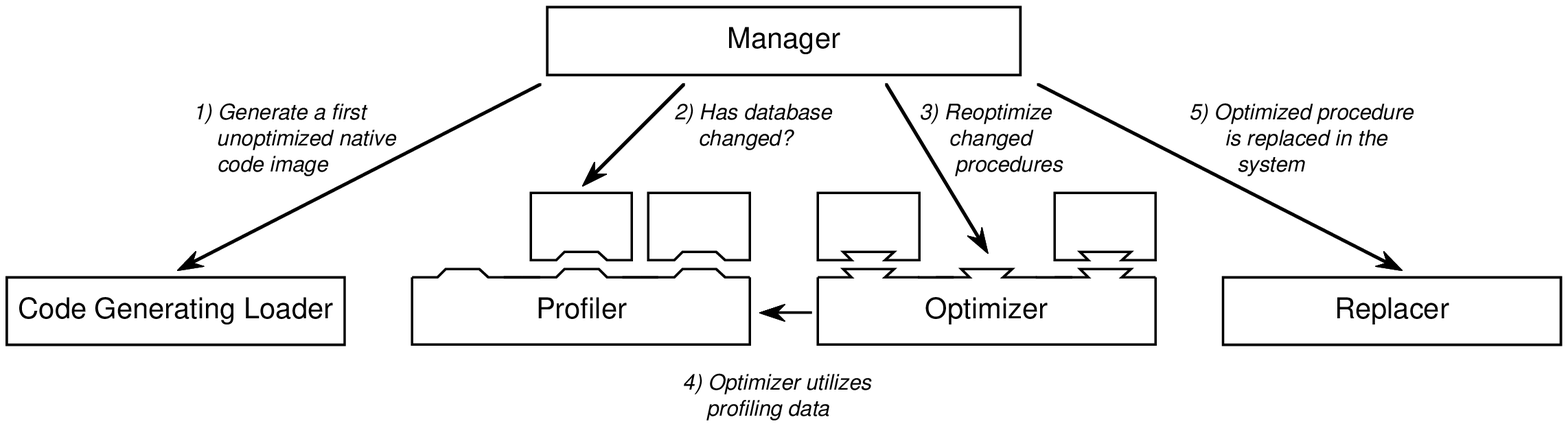}}
\caption{General Architectural Overview}
\end{figure}

The five constituents of our system interact as follows: When the user first launches
an application, the \emph{code-generating loader} translates the representation\footnote{Our particular 
implementation uses the Slim Binary representation \cite{FK97}, but the architecture presented here does not 
depend on this fact.  The same architecture could also be used with programs represented as class files for 
the Java Virtual Machine, or even with native code for some specific processor.  This would merely have an 
effect on the pre-processing effort required to extract information relevant to the optimizer, such as control 
flow and data flow information.} that programs are transported in into a sequence of native machine 
instructions.  Because this is an interactive process (the user is waiting), and because many worthwhile code 
optimizations are extremely time-intensive, the code-generating loader doesn't optimize much but instead 
concentrates on simply getting the program to run as quickly as possible.

Once that the application program has begun to execute, the \emph{profiler} starts collecting information about 
its behavior.  This information is later used to guide optimizations.  Examples of the kinds of information 
collected by the profiler are the call-frequencies of individual procedures, statistics on how variables and 
parameters are accessed, and a catalog of which instructions stall due to misses in the data cache.  The 
profiler runs continuously at all times.  It has an extensible structure that can support a wide spectrum of 
profiling techniques and can be augmented as required by the plug-and-play addition of appropriate \emph{
profiling components}, such as instrumenting profilers and sampling profilers.

The \emph{system manager} executes a low priority thread that uses application idle time to optimize the 
already running software in the background.  It repeatedly queries the profiling database to examine whether 
the characteristics of the system's behavior have changed, and for which procedures they have changed.  Based 
on this information, the system manager builds a list of procedures to potentially optimize.  The optimization 
candidates in this list aren't all equally well suited for optimizations---optimizing some procedures might be 
more profitable than optimizing others.  Therefore, the system manager additionally queries the optimization 
manager for a rough estimate of the profitability of optimizing each individual procedure.  This information 
is used to sort the candidate list according to optimization priorities.

The system manager then invokes the \emph{optimizer} on each procedure, in the order in which they appear in 
the candidate list.  The optimizer is driven by the profiling database and can concentrate its efforts on 
those parts of the system that are most critical and most beneficial to optimize.  This is in contrast to 
traditional code optimizers that apply optimizations uniformly to the entire code base.  Similar to the 
profiler, the optimizer is configurable by adding, removing, or replacing \emph{optimization components}.

Finally, after the optimizer has completed its work, the \emph{replacer} hot-swaps the currently executing code 
image against the newly generated, optimized version.  This process requires updating interprocedural and 
intermodular dependencies and, in some cases, undoing previous optimizations.
Listing 1 illustrates the functionality of the system manager as pseudo-code.

\begin{acmlisting}[htb]
\scriptsize
{\tt 
\hspace*{2em} PROCEDURE SystemManager(); \newline
\hspace*{2em} \hspace*{0.2 cm} VAR  \newline
\hspace*{2em} \hspace*{0.2 cm} \hspace*{0.2 cm} P: Procedure; \newline
\hspace*{2em} \hspace*{0.2 cm} \hspace*{0.2 cm} estimate, sleepTime: LONGINT; \newline
\hspace*{2em} \hspace*{0.2 cm} \hspace*{0.2 cm} oldAgeTime, oldSimilarityTime: System.Time; \newline
\hspace*{2em} BEGIN \newline
\hspace*{2em} \hspace*{0.2 cm} sleepTime $\leftarrow$ 1; oldAgeTime $\leftarrow$ 0; oldSimilarityTime $\leftarrow$ 0; \newline
\hspace*{2em} \hspace*{0.2 cm} LOOP \newline
\hspace*{2em} \hspace*{0.2 cm} \hspace*{0.2 cm} Threads.Sleep(Thread.This(), sleepTime); \newline
 \newline
\hspace*{2em} \hspace*{0.2 cm} \hspace*{0.2 cm} {\it (* periodically age profiling data *)} \newline
\hspace*{2em} \hspace*{0.2 cm} \hspace*{0.2 cm} IF System.GetTime() > oldAgeTime + AgeSleepTime THEN \newline
\hspace*{2em} \hspace*{0.2 cm} \hspace*{0.2 cm} \hspace*{0.2 cm} ProfileSystem.Age() \newline
\hspace*{2em} \hspace*{0.2 cm} \hspace*{0.2 cm} \hspace*{0.2 cm} oldAgeTime $\leftarrow$ System.GetTime(); \newline
\hspace*{2em} \hspace*{0.2 cm} \hspace*{0.2 cm} END; \newline
 \newline
\hspace*{2em} \hspace*{0.2 cm} \hspace*{0.2 cm} {\it (* periodically check whether profiling data has changed *) }\newline
\hspace*{2em} \hspace*{0.2 cm} \hspace*{0.2 cm} IF System.GetTime() > oldSimilarityTime + SimilaritySleepTime THEN \newline
\hspace*{2em} \hspace*{0.2 cm} \hspace*{0.2 cm} \hspace*{0.2 cm} FORALL procedures P DO  \newline
\hspace*{2em} \hspace*{0.2 cm} \hspace*{0.2 cm} \hspace*{0.2 cm} \hspace*{0.2 cm} IF NOT ProfilingSystem.StableProc(P) THEN  \newline
\hspace*{2em} \hspace*{0.2 cm} \hspace*{0.2 cm} \hspace*{0.2 cm} \hspace*{0.2 cm} \hspace*{0.2 cm} {\it (*profiling data for P is not stable: estimate the  \newline
\hspace*{2em} \hspace*{0.2 cm} \hspace*{0.2 cm} \hspace*{0.2 cm} \hspace*{0.2 cm} \hspace*{0.2 cm} \hspace*{0.2 cm} profitability of reoptimizing P *)} \newline
\hspace*{2em} \hspace*{0.2 cm} \hspace*{0.2 cm} \hspace*{0.2 cm} \hspace*{0.2 cm} \hspace*{0.2 cm} estimate $\leftarrow$ OptimizationSystem.Estimate(P); \newline
\hspace*{2em} \hspace*{0.2 cm} \hspace*{0.2 cm} \hspace*{0.2 cm} \hspace*{0.2 cm} \hspace*{0.2 cm} IF estimate > 5\% THEN \newline
\hspace*{2em} \hspace*{0.2 cm} \hspace*{0.2 cm} \hspace*{0.2 cm} \hspace*{0.2 cm} \hspace*{0.2 cm} \hspace*{0.2 cm} {\it (*add P to the list of procedures to be optimized; in this list,  \newline
\hspace*{2em} \hspace*{0.2 cm} \hspace*{0.2 cm} \hspace*{0.2 cm} \hspace*{0.2 cm} \hspace*{0.2 cm} \hspace*{0.2 cm} \hspace*{0.2 cm} procedures are sorted by descending profitability estimate *)} \newline
\hspace*{2em} \hspace*{0.2 cm} \hspace*{0.2 cm} \hspace*{0.2 cm} \hspace*{0.2 cm} \hspace*{0.2 cm} \hspace*{0.2 cm} OptimizationSystem.AddProcedureToSchedule(P, estimate) \newline
\hspace*{2em} \hspace*{0.2 cm} \hspace*{0.2 cm} \hspace*{0.2 cm} \hspace*{0.2 cm} \hspace*{0.2 cm} END \newline
\hspace*{2em} \hspace*{0.2 cm} \hspace*{0.2 cm} \hspace*{0.2 cm} \hspace*{0.2 cm} END \newline
\hspace*{2em} \hspace*{0.2 cm} \hspace*{0.2 cm} \hspace*{0.2 cm} END; \newline
\hspace*{2em} \hspace*{0.2 cm} \hspace*{0.2 cm} \hspace*{0.2 cm} oldSimilarityTime $\leftarrow$ System.GetTime(); \newline
\hspace*{2em} \hspace*{0.2 cm} \hspace*{0.2 cm} END; \newline
\hspace*{2em}  \newline
\hspace*{2em} \hspace*{0.2 cm} \hspace*{0.2 cm} {\it (* pick the next procedure to optimize *)} \newline
\hspace*{2em} \hspace*{0.2 cm} \hspace*{0.2 cm} P $\leftarrow$ OptimizationSystem.GetNextScheduledProcedure(); \newline
\hspace*{2em} \hspace*{0.2 cm} \hspace*{0.2 cm} IF P $\neq$ NIL THEN  \newline
\hspace*{2em} \hspace*{0.2 cm} \hspace*{0.2 cm} \hspace*{0.2 cm} OptimizationSystem.Optimize(P); \newline
\hspace*{2em} \hspace*{0.2 cm} \hspace*{0.2 cm} \hspace*{0.2 cm} Replacer.Replace(P); \newline
\hspace*{2em} \hspace*{0.2 cm} \hspace*{0.2 cm} \hspace*{0.2 cm} sleepTime $\leftarrow$ 1 \newline
\hspace*{2em} \hspace*{0.2 cm} \hspace*{0.2 cm} ELSE \newline
\hspace*{2em} \hspace*{0.2 cm} \hspace*{0.2 cm} \hspace*{0.2 cm} {\it (* nothing to optimize: adjust the sleep time that passes until the  \newline
\hspace*{2em} \hspace*{0.2 cm} \hspace*{0.2 cm} \hspace*{0.2 cm} \hspace*{0.2 cm} system manager reconsiders optimizations *) }\newline
\hspace*{2em} \hspace*{0.2 cm} \hspace*{0.2 cm} \hspace*{0.2 cm} sleepTime $\leftarrow$ sleepTime * 2; \newline
\hspace*{2em} \hspace*{0.2 cm} \hspace*{0.2 cm} \hspace*{0.2 cm} IF sleepTime > AgeSleepTime THEN sleepTime $\leftarrow$ AgeSleepTime END; \newline
\hspace*{2em} \hspace*{0.2 cm} \hspace*{0.2 cm} END \newline
\hspace*{2em} \hspace*{0.2 cm} END \newline
\hspace*{2em} END SystemManager; \newline
} 

\normalsize
\caption{System Manager}
\end{acmlisting}

The remainder of this paper concentrates on design aspects of the individual components
of our system---that is the optimizer, the profiler, and the replacer. The design
of code-generating loaders and other just-in-time compilers has already been documented
elsewhere \cite{DS84,Fra94,Hol94,HU96,ACL+98} and our implementation provides no meaningful new insights on 
this topic.

\section{The Optimizer Subsystem} 
The optimizer is composed of three main parts: the optimization manager, a history
database, and a set of optimization components that can be dynamically added, removed,
and exchanged. Figure 2 presents a schematic overview of how these parts interact.
The various services offered by the optimizer operate on the program being optimized
at the level of individual procedures. However, the fact that the optimizer operates
on a program using a granularity of one procedure at a time does not imply that it
cannot perform interprocedural optimizations---it merely represents an architectural
decision. Our implementation preserves cross-procedural state in the history database
and in individual optimization components.

\begin{figure}[htb]
\epsfxsize=4.5 in
\centerline{\epsffile{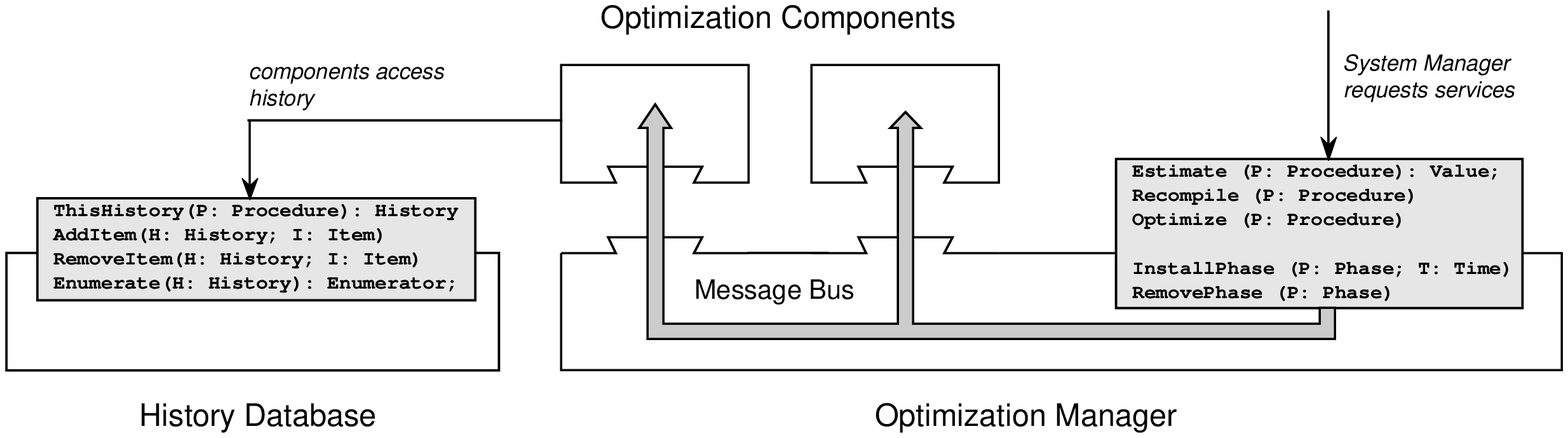}}
\caption{Schematic View of the Optimizer}
\end{figure}

The \emph{optimization manager} handles all requests from the system manager and coordinates the optimization 
process.  Optimizations are organized as a sequence of \emph{phases} that are executed sequentially.  The 
various phases operate on a common intermediate representation of the program, \emph{guarded single assignment 
form} (GSA) \cite{Bra95}, a variant of SSA \cite{CFR+91}.  The intermediate GSA representation is not cached 
across separate invocations of the optimizer but generated afresh from the software transportation 
format\footnote{i.e., in the case of our implementation, the aforementioned Slim Binary format.} each time 
that a new optimization cycle commences---since the unit of optimization is the procedure, this keeps memory 
consumption within reasonable limits.  Hence, the first phase of the optimizer generates GSA for a procedure 
of the program being optimized, while each subsequent phase retrieves this intermediate representation, 
performs a specific task, and then returns a possibly modified version of the procedure in the same 
intermediate format.  If a look at actual profiling data suggests that an optimization is not profitable, the 
intermediate representation remains unmodified.  The specific tasks that individual phases perform correspond 
to individual code optimizations such as dead code elimination, common subexpression elimination, and register 
allocation.

An \emph{optimization component} is a container that encapsulates the implementation of one or more 
optimization phases.  In most cases, each component implements exactly one phase.  As 
discussed in the introduction, plug-and-play customizability makes it possible to achieve a perfect match 
between user software and underlying hardware platform without requiring global updates of either application 
programs or the run-time system.  Instead, for each new member of a processor family, the hardware 
manufacturer merely needs to supply the specific components that perform optimizations tailored to 
characteristics in which the new processor differs from the generic representative of the family.  As an 
example, there would be a unique instruction-scheduling component for each processor model.  As a further 
example, a component supporting MMX instructions would map specific library calls and possibly further code 
patterns directly into multimedia instructions emitted in-line.

The \emph{history database} is a repository that records the set of optimizations that have been performed on 
individual procedures.  It is used for bookkeeping and for coordinating code optimizations.  Since different 
optimization techniques may have conflicting goals (e.g., ``decrease code-size to reduce misses in the 
instruction cache'' vs.  ``unroll loops to reduce misses in the data cache''), an optimization phase may 
consult the history database to determine whether its technique interferes with previously applied 
optimizations.  This is important because phases are independent of each other; each phase is only aware of 
which other phases have executed previously in the current optimization schedule for the procedure under 
consideration, and completely unaware about phases that might follow further downstream.  The history database 
is kept in memory and its contents are volatile: history information is not preserved across cold-starts of 
our system, and even while the system is running, all global state information is ``aged'' periodically (see 
below).

\subsection*{Component Interaction in the Optimizer} 
The key to flexibility in our solution is that new phases can be registered at the
optimization manager, can be removed from the optimizer, and can even replace other
phases without affecting the remainder of the run-time system. This is achieved by
letting the optimization manager communicate with individual phases via an open central
message bus, rather than hard-coding component interfaces. When the optimization
manager receives a request from the system manager, it translates the request into
a sequence of messages and distributes them to the phases within installed optimization
components. Each optimization phase has to conform to the message protocol depicted
in Figure 3. In the following, we explain the semantics attached to the individual
messages.

\begin{figure}[htb]
\epsfxsize=3 in
\centerline{\epsffile{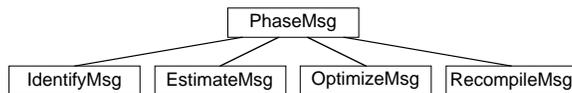}}
\caption{Optimizer Message Protocol}
\end{figure}

When the profiler subsystem detects a substantial change in the behavior of a certain
procedure, the system manager invokes the optimizer's {\tt Estimate()} service. The
{\tt Estimate()} service assesses the profitability of applying further optimizations
to the procedure. Based on this assessment, the system manager then decides whether
or not to actually optimize the procedure. Since the estimate is used to eliminate
unprofitable optimization candidates, it has to be computed efficiently---at least
in relation to the time that would be spent optimizing those candidates. Consequently,
this computation is based on simple heuristics without actually looking ``into''
the optimization candidate itself. This also means that the estimate can be computed
without first having to generate GSA.

The {\tt Estimate()} service is implemented by sending an {\tt EstimateMsg} to each
installed optimization phase. Each phase is responsible for computing the estimated
speedup that would result if the associated optimization were added to the already
existing optimization schedule for a given procedure. Hence, if the associated optimization
is currently already performed on the procedure, no additional speedup can be expected
and a value of zero is returned. Otherwise, a simple heuristic is used to compute
the profitability of optimizing the procedure. The heuristic is based both on a hard-coded
average speedup (e.g., 5\% speedup for data prefetching vs. 20\% speedup for common
subexpression elimination) and actual speedups measured for previous applications
of the same optimization (to this and other procedures). The total speedup estimate for
a procedure is then derived by computing the sum of the speedup estimates of all
optimization phases that are present in the system.

Once the system manager has decided which procedures to optimize, it invokes the
{\tt Optimize()} service for each optimization candidate. The {\tt Optimize()} service
creates a newly optimized version of a given procedure from scratch. To begin with,
a GSA representation of the procedure is generated from the software transportation
format (e.g., the original ``object file'' or an in-memory cache). Then, each installed
optimization phase is sent an {\tt OptimizeMsg}, instructing it to apply its respective
modifications to the procedure's GSA representation (the order in which individual
phases execute is discussed below). Finally, the optimized GSA representation is
transcribed into native code and handed over to the replacer.

Upon receiving an {\tt OptimizeMsg}, an optimization phase first needs to re-evaluate
whether or not it would be profitable to perform its associated task. This is because
the original estimate based upon which the optimization manager decided to apply
this optimization was founded on inaccurate low-cost heuristics, whereas at this
point, the full GSA representation and up-to-the-minute profiling data are available.
This makes a much better estimate of the anticipated speedup possible. For example,
a phase performing loop unrolling might have an estimated speedup of 10\%. However,
after looking at the loops in question, the optimization phase might determine that
they exhibit too little parallelism and hence aren't profitable after all.

If the optimization had previously been applied to the given procedure, its benefit
is re-evaluated on the basis of actual performance data. If the past speedup was
negative or insignificant, the optimization is marked as non-profitable in the history
database and is dropped from future iterations. Otherwise, it is applied again. If
the particular code optimization had previously not been applied to the procedure,
the optimization phase examines profiling data and decides whether to apply it or
not (for example, based on whether a certain profiling counter exceeds a certain
threshold). If it decides to apply the optimization, it is added to the history database
and is performed.

In contrast to the {\tt Optimize()} service that applies optimizations \emph{optimistically}, the {\tt 
Recompile()} service optimizes procedures \emph{pessimistically}.  Neither does it consider new optimizations 
nor does it remove unsuccessful old ones.  It only re-performs the set of optimizations recorded in the 
history database, excluding optimizations previously determined to be non-profitable.  This service is 
particularly useful for \emph{de-optimizations}, a case in which optimizations have to be selectively undone.  
We already gave an example where this is useful, in the case where a statically bound method is overridden in 
a dynamically loaded extension.  In such a situation, the previous optimization must be undone.  This is 
achieved by removing it from the history database and calling the {\tt Recompile()} service for all affected 
methods.  The {\tt Recompile()} service is implemented by sending a {\tt RecompileMsg} to all installed 
optimization phases.

Finally, the {\tt IdentifyMsg} is sent to an optimization phase to request meta-information,
such as its name and when it wishes to be executed during the optimization process.
Our architecture does not attempt to solve the general \emph{phase-ordering problem} of compiler construction 
\cite{CC95}.  In our current implementation, the various phases ``know'' their relative place in an ideal 
schedule, under the implied assumption that this can somehow be coordinated by extension providers.  A 
newly loaded optimization component can inspect the set of already present phases and then install its 
constituent phases immediately before or immediately after any already existing phase.  A single component can 
contain several phases that execute at different points in the time-line, and copies of the same phase can be 
inserted at multiple points in the time-line.  Still, we acknowledge that this solution is 
sub-optimal, which is why the issue is revisited under the heading of ``open questions'' below.

\section{The Profiler Subsystem} 
It has been almost thirty years since it was first realized that code quality could
be improved by using feedback information (i.e., execution profiles) to guide optimizations
\cite{Ing71}.  As processor complexity has been rising, the number of optimization decisions that a compiler must 
make has grown accordingly.  Unfortunately, making the wrong optimization choices can seriously affect runtime 
performance.  Access to profiling information makes it possible to base optimization decisions (such as which 
procedures to inline, which execution paths to favor during scheduling, and which variables to spill to 
memory) on actual measured performance data rather than on imprecise (and often ad-hoc) heuristics.

The most accurate profile of a program's execution can be obtained by \emph{simulating} the processor under 
consideration as well as the relevant parts of the memory hierarchy at the gate level.  However, this approach 
is hardly feasible in a system that aims to respond to changing user needs almost in real time.  Hence, our 
approach considers only profiling techniques that are applicable in real-time situations.  These fall into the 
three main categories of \emph{instrumentation} \cite{MIPS90,BL94}, \emph{sampling} \cite{ABD+97,ZWG+97}, and \emph{
hardware-based solutions} \cite{DHW+97,CMH96}.

The last of these is obviously the most desirable and will become increasingly important
in future microprocessors. As an example, while the 601 and 603 models of the PowerPC
processor family \cite{Mot97} do not provide built-in profiling support, the PowerPC 604 is now equipped with a 
performance monitor.  This performance monitor includes two 32-bit hardware counters that facilitate 
monitoring detailed events during execution, such as instruction dispatches, instruction cycles, misses in the 
cache, and load/store miss-latencies.  The PowerPC 604e even includes four counters with augmented 
functionality.

For the foreseeable future, however, system builders will have to accept the fact
that hardware profiling support is inadequate. Consequently, one has to rely on either
or both of the other two techniques. Neither of them is fully appropriate for capturing
the entire spectrum of profiling needs. On the one hand, instrumentation is well
suited for generating exact path profiles \cite{BL96}; sampling techniques fail in this task because temporal 
information is lost in the statistical process.  On the other hand, a sampling profiler can quite 
accurately pin down the set of instructions that miss in the data cache (the likelihood of the program counter 
hitting such an instruction is higher than the likelihood of hitting another instruction).  An instrumenting 
profiler cannot usually determine whether an instruction missed in the cache, unless it is assisted by special 
hardware counters.

As a consequence, a sound profiling infrastructure has to support both sampling and
instrumenting profilers, and be extensible to hardware-based profiling as it becomes
available. This points toward an architecture that is surprisingly similar to that
of the extensible optimizer introduced above. In our implementation, the profiling
subsystem is composed of a profiling manager and a set of ``plug-in'' profiling components,
as presented in Figure 4.

\begin{figure}[htb]
\epsfxsize=4.5 in
\centerline{\epsffile{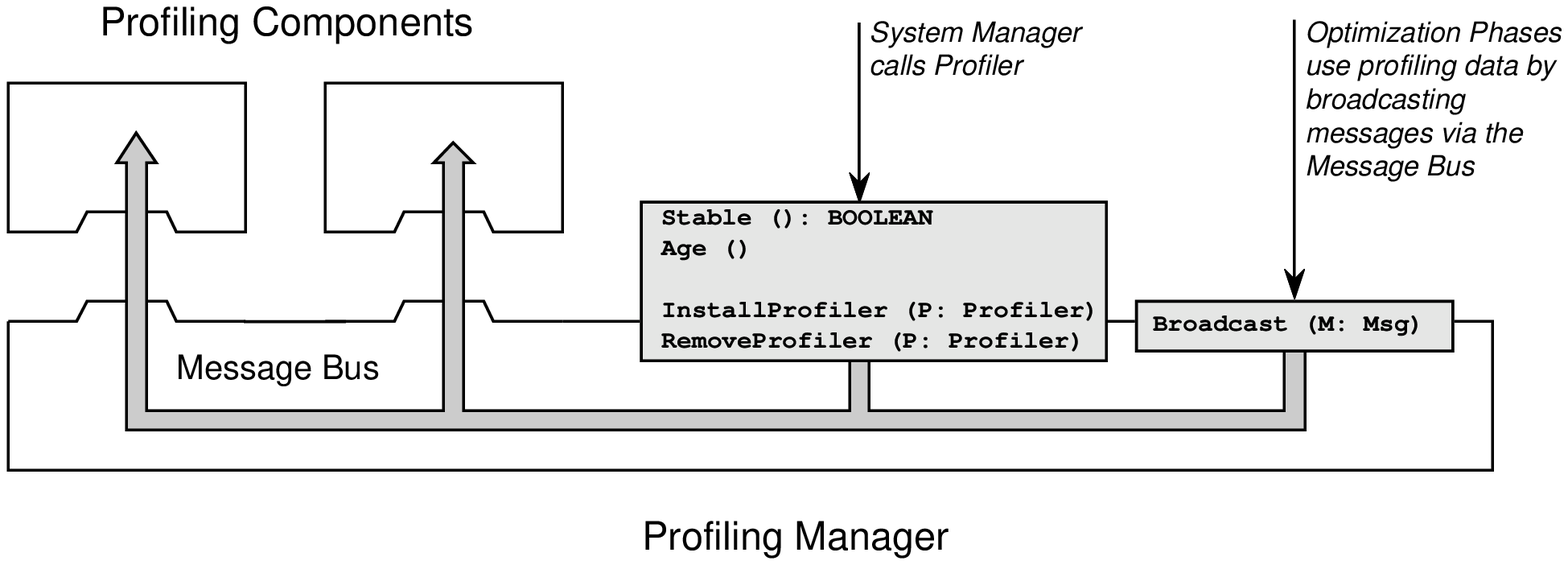}}
\caption{Schematic View of the Profiler}
\end{figure}

Just as in the optimizer, communication between the profiling manager and the installed
profiling components is achieved by a broadcast mechanism via a message bus. Optimization
components request profiling information by sending messages to the profiling subsystem,
which in turn delegates these requests to the appropriate \emph{profiling component(s)}.  As far as an 
optimization component is concerned, it is not relevant which profiling component actually processes its 
requests, as long as there is a component that does.  This is the key to the evolvability offered by our 
solution. For example, when suitable hardware support for profiling becomes available, sampling and 
instrumenting profilers can be replaced by hardware-assisted ones simply by providing an appropriate profiling 
component that maps profiling requests directly onto the appropriate hardware counters.

Extensibility of the profiler in such a plug-and-play fashion also makes it possible
to meet unanticipated requirements of future optimization phases: Suppose that a
new plug-in optimization would require a specific kind of performance information
that is not provided by the default profiling system. This problem can be solved
easily by pairing the new optimization component with a dedicated profiling component
that supplies it with the needed data.

Unlike our optimizer subsystem, the profiler doesn't provide a centralized data\-base.
This is because the kinds of data that the various profiling components collect and
store are highly divergent in nature, and the availability of hardware-assisted profiling
would eventually lead to an additional overhead for keeping the database in synch
with the hardware counters. In our architecture, individual profiling components
are autonomous and store their profiling data separately; in the current implementation,
all of this information is kept entirely in memory. As is elaborated in the following,
the profiling subsystem provides a centralized service for periodically and synchronously
aging the information in this distributed database.

\subsection*{Component Interaction in the Profiler} 
A profiling component needs to adhere to a particular message protocol that is illustrated
in Figure 5. In the following, we give an overview of the services that are associated
with these messages.

\begin{figure}[htb]
\epsfxsize=3.5 in
\centerline{\epsffile{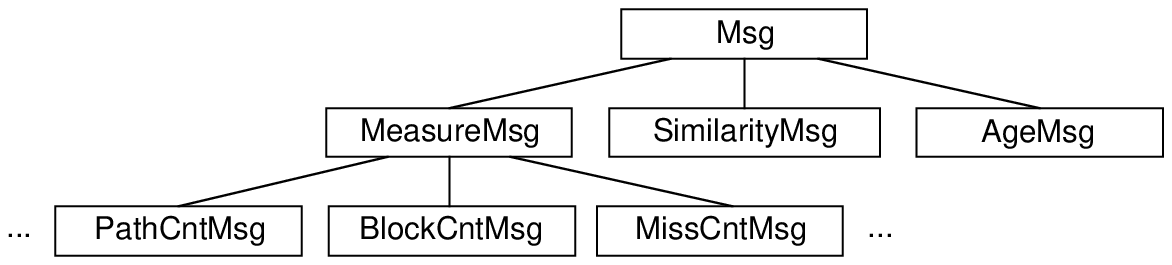}}
\caption{Profiler Message Protocol}
\end{figure}

The {\tt AgeMsg} is broadcast to all profiling components when the system manager
invokes the {\tt Age()} service. This is done periodically to adjust and reduce the
relevance of older profiling data. The implementation of aging is left to each individual
profiling component, with exponential decay or linear decay being possible models.

The system manager also periodically checks whether the system's behavior has shifted
over time by calling the profiler subsystem's {\tt Stable()} service. The {\tt Stable()}
service returns false if the profiling data has substantially changed since the last
{\tt Age()} request---otherwise it returns true. The profiler manager reacts to a
{\tt Stable()} request by broadcasting a {\tt SimilarityMsg} to all profiling components.
Individual profiling components react to this message by computing a similarity measure
that reflects the degree of change in their profiling data for a given period of
time. Section 5 discusses this similarity measure in more detail. 

The primary means for optimization phases to communicate with the profiler is the
message broadcast mechanism. Whenever an optimization phase requires profiling information,
it creates an instance of a particular message and passes it to the profiling manager's
{\tt Broa}{\tt d}{\tt cast()} service. In turn, the {\tt Broadcast()} service distributes
the message to all the installed profiling components. For each profiling event that
needs to be monitored, a new message type is derived\footnote{Note that this is
an open-ended interface: the range of profiling events to be monitored cannot be
determined in advance, as future optimization components might have requirements
that simply cannot be anticipated. The way to solve this problem is by encapsulating
the requests themselves as a ``message objects'' derived from a common superclass.
This is also known as the \emph{Command (233)} design pattern \cite{GHJV95}.} from the common {\tt 
MeasureMsg}.  Examples include messages for measuring execution path frequencies ({\tt PathCntMsg}) and 
data-cache miss rates ({\tt MissCntMsg}).  If a particular profiling component receives such a message and 
provides a service for that profiling event, it takes appropriate actions, otherwise the message is ignored.

Details about the request itself are encoded in the message and may contain one of
three different request codes: (1) An optimization component may signal interest
in a particular profiling event (e.g., ``measure the execution frequency of path
14''), in which case the profiling component sets up auxiliary data structures to
store the corresponding profiling data. It may also direct an optimization component
to insert instrumentation code at the code location under consideration. (2) An optimization
component may request profiling data for a particular event type (e.g., ``return
the current execution frequency for path 14''). And (3), an optimization component
may signal that a specific set of profiling information is no longer needed (e.g.,
``the frequency of path 14 does not have to be measured any longer''). This is usually
the case after optimizations have been performed. The profiling component then de-allocates
auxiliary data structures and instructs an optimization component to remove previously
installed instrumentation code.

Another significant property of our architecture is that new profiling components
can easily be composed out of existing components, both vertically and horizontally.
New components can share existing functionality via message forwarding. As an example,
we can construct a basic block profiler on top of a path profiler. Whenever the basic
block profiler receives a {\tt BlockCntMsg}, it creates and re-broadcasts a new {\tt PathCntMsg}
for each path that crosses the specified basic block. The basic block count is then
computed by summing up the path counts for the individual paths. Neither does the
new event profiler have to store additional data nor does it have to know implementation
details of the path profiler.

The component-oriented architecture also facilitates a particularly elegant solution
to the problem of constructing \emph{instrumenting profilers}.  Instrumentation involves modifying the actual 
machine code that is executed by the processor.  The traditional approach has been to use binary rewriting 
tools \cite{ES94} that insert instrumentation only after the final code images of programs have already been 
generated.  Our solution, on the other hand, consists of structuring the instrumenting profiler as a closely 
coupled pair consisting of a profiling component and a corresponding optimization phase that inserts 
instrumentation code directly into the GSA representation of a procedure (Figure 6).  If the instrumentation 
phase comes early enough in the optimization schedule, this has the beneficial effect that profiling 
instructions are automatically optimized in the context of the procedure, a very important advantage if 
low-overhead continuous profiling is an objective.  Also note that the optimizer doesn't need to be modified 
in any way in order to support instrumentation---this is a natural capability provided by its component 
architecture.

\begin{figure}[htb]
\epsfxsize=3.5 in
\centerline{\epsffile{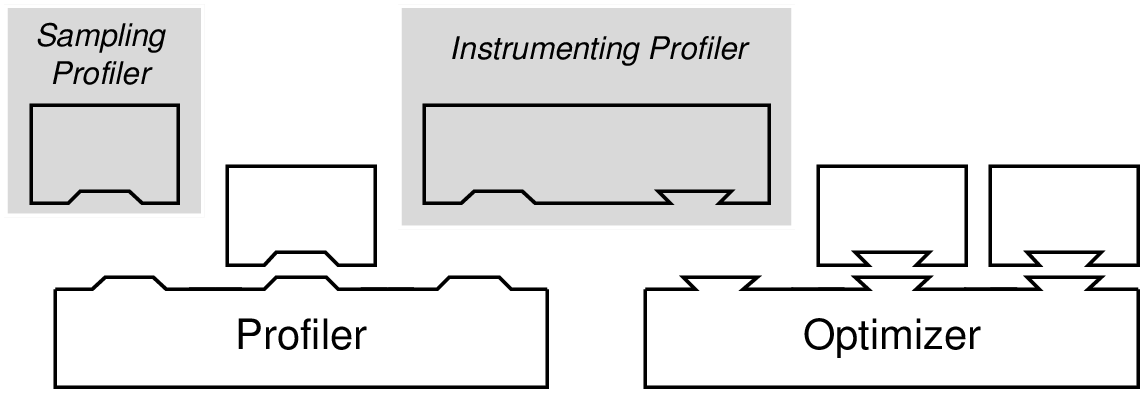}}
\caption{Implementation of Sampling Profilers vs. Instrumenting Profilers}
\end{figure}

\section{Computing the Similarity of Profiling Data} 
One of the essential problems when performing optimizations at runtime is to decide
\emph{when} to optimize and \emph{what} to optimize.  Optimizing too little does not greatly improve runtime 
performance, optimizing too aggressively might lead to a situation in which the effort invested into code 
optimization is never fully recouped by faster-running application code \cite{HU96}.

This section addresses the first question, \emph{when} to optimize.  In our system, optimizations are initially 
performed when a program has been launched and enough profiling data has been gathered.  Additionally, 
optimizations are reconsidered whenever the footprint of the profiling data changes substantially, i.e., when 
the user's behavior has shifted noticeably.  In such a case, earlier optimizations may no longer align well 
with the current use of the system, and optimum performance may be restored only by performing optimization 
all over again.

In order to detect substantial changes in the user's behavior, we define a similarity
measure $S$ that reflects the degree of change of profiling data between two consecutive
time steps $t-1$ and $t$. Each profiling component $P$ logs $n$ distinct values (such
as a path counter or a basic block counter) that we represent as an $n$-dimensional
vector $\vec{p}$, and is required to log these profiling values for at least
the last two time steps. The similarity measure $S(P)$ can then be expressed as a
function $S:P\rightarrow[0..1]$ that compares the captured data at time step $t-1$
(i.e., $\vec{p}_{t-1}$) with the captured data at time step $t$ (i.e., $\vec{p}_{t}$).
It returns a similarity value in the range $[0..1]$, whereas 0 denotes complete dissimilarity
and 1 denotes complete data equivalence.

We first try to define $S(P)$ as a function that computes the geometric angle $\alpha$
between $\vec{p}_{t-1}$ and $\vec{p}_{t}$:

\belowdisplayshortskip=12pt
$$ \alpha = \arccos{\frac{\vec{p}_{t-1} \cdot \vec{p}_{t}}{\left| \vec{p}_{t-1} \right| \left| \vec{p}_{t} \right|}} $$

This term has the advantageous property that it is independent of the time difference
between $t-1$ and $t$ since it measures the angle between the two vectors only and
disregards the length of the vectors. However, it is not defined in the situation
where  $\vec{p}_{t-1}=\vec{0}$ and $\vec{p}_{t}=\vec{0}$. This
is the case when the profiling database is first set up and initialized for a newly
loaded application. To eliminate this problem, we adjust $\alpha$ by adding 1 to
the denominator of the term. For simplicity reasons, we also remove the $arccos$
function. The remaining term is still continuously descending and allows us to set
a threshold for reconsidering new optimizations:

\belowdisplayshortskip=12pt
$$ \alpha = \frac{\vec{p}_{t-1} \cdot \vec{p}_{t}}{\left| \vec{p}_{t-1} \right| \left| \vec{p}_{t} \right| + 1} $$

However, this function has further undesirable properties: It is very sensitive to
small changes for short and low dimensional vectors  $\vec{p}$. For example,
if we measure the execution frequency of two paths, both paths have been executed
once at time step $t-1$ ($\vec{p}_{t-1}=(1,1)$), and one path is
executed once more between $t-1$ and $t$ ($\vec{p}_{t}=(2,1)$), the
resulting $\alpha$ suggests a considerable change in the profiling database---which
of course is true, but an absolute change by only 1 should clearly not trigger a
reoptimization. An optimal function should therefore disregard changes smaller than
a given threshold. To achieve this, we define a second term $\beta$ that reflects
the absolute size of the change: 

\belowdisplayshortskip=12pt
$$ \beta = \frac{\left| \vec{p}_{t} - \vec{p}_{t-1} \right| }{\sqrt{n}} $$

Note that this term is independent of the dimension of $\vec{p}$ since the absolute
change is divided by $\sqrt{n}$ (the unit vector of dimension $n$ has length
$\sqrt{n}$).  We can now redefine the similarity function $S(P)$ as a combination of the angular component 
$\alpha$ and the length component $\beta$:

\belowdisplayshortskip=12pt
$$ S(P) = e^{- {\left( \frac{\beta}{c} \right)}^{k}}(1 - \alpha) + \alpha $$

As illustrated in Figure 7, for large vectors, the function still returns the geometric
angle between the two vectors $\vec{p}_{t-1}$ and $\vec{p}_{t}$ since
it strives towards $\alpha$. For small vectors, however, the function strives towards
1 and is less sensitive to small changes as a result. It even completely disregards
changes smaller than $c$---the constant $c$ in the term was chosen to approximate
the turning point of the function. By appropriately setting $c$, we can adjust the
threshold above which changes in profiling data gets reflected in the similarity
measure $S$ (e.g., a procedure is only optimized if it has been executed at least
100 times in the last time period). Similarly to $c$, the constant $k$ can be used
to modify the slope of the function. We have found that a value of 8 performs quite
well in practice.

\begin{figure}[htb]
\epsfxsize=4.5 in
\centerline{\epsffile{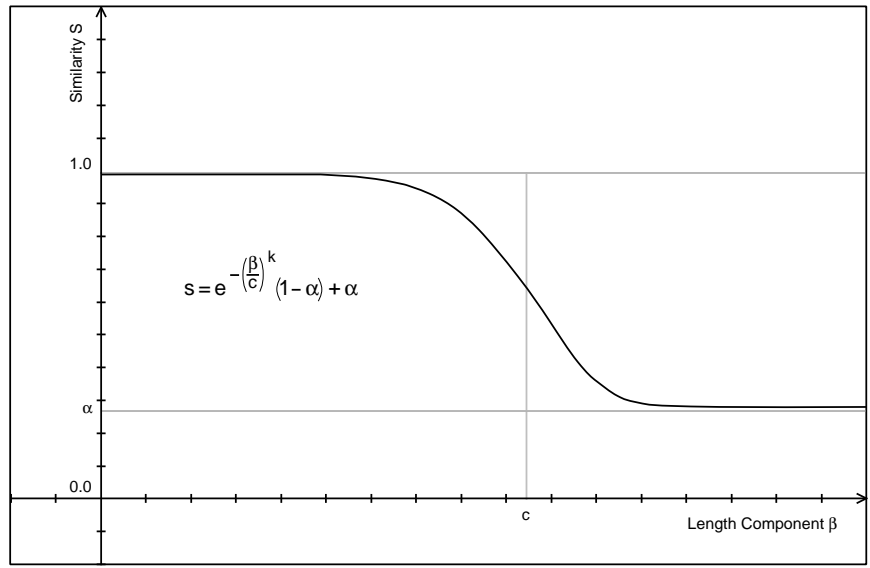}}
\caption{Similarity Function}
\end{figure}

One more problem is still remaining, though: The function $S(P)$ always returns 1
for vectors of dimension 1. This can be circumvented elegantly by adding an additional
component $n$ to both $\vec{p}_{t-1}$ and $\vec{p}_{t}$ with $\vec{p}_{t-1,n}=m$,
$\vec{p}_{t,n}=m$, and $m=max(\vec{p}_{t-1,0},\dots, \vec{p}_{t-1,n-1},\vec{p}_{t,0},\dots, \vec{p}_{t,n-1})$.

In practice, we say that profiles have not changed as long as $S(P)=1.0$. We reconsider
existing optimizations when $S(P)<0.95$.

\section{Replacing Code} 
In this section, we address the questions of \emph{how} code is replaced and \emph{when} code is replaced after 
optimization.  There are several different methods for replacing the code image of a procedure by a newly 
optimized image of the same procedure.

Most straightforwardly, we can simply overwrite the old code image by a new one \emph{in situ}, preserving the 
existing entry point.  This has the apparent advantage that no branch instructions terminating at the 
procedure's entry point need to be updated.  Unfortunately, this only works well as long as the new image is 
at most as large as the old image, which is not very likely given that many optimizations increase the code 
size in exchange of greater speed.  Examples of such optimizations include loop unrolling, prefetching, 
loop tiling, and procedure inlining.  The code size problem can be alleviated by reserving space ``between'' 
procedures in the code image, but this leads to memory fragmentation.  Moreover, overwriting existing code 
images causes problems if the procedure being replaced is active at the time of replacement.

Consequently, the new code image is preferably stored in a new, separate memory region.
This preserves the old code image but requires changing interprocedural dependencies---branches
to the old image need to be replaced by branches to the new image. Similarly, and
less simply, the contents of \emph{procedure variables} pointing to a modified procedure need to be updated, 
and assignments of the affected procedure to procedure variables need to be replaced by corresponding 
assignments of the procedure's entry address in the new code image.

These updates can be performed either \emph{immediately} or \emph{lazily}.  Updating dependencies lazily 
involves replacing the beginning of the old version by a code stub that not only redirects any eventual caller 
to the new location, but that also modifies the calling location so that the new destination address is 
reached directly upon future calls.  Since the stub that performs all of these actions is often rather large 
itself, a second indirection is usually employed.  This is illustrated in Figure 8: The first three 
instructions of the old code image are replaced with code that loads the address of the new code image.  This 
address is then passed to the procedure {\tt D}{\tt i}{\tt vert\_Call}, which replaces the original call of 
the old procedure by a call of the new procedure.  After the modification, control passes to the new 
procedure.

\begin{figure}[htb]
\epsfxsize=4.5 in
\centerline{\epsffile{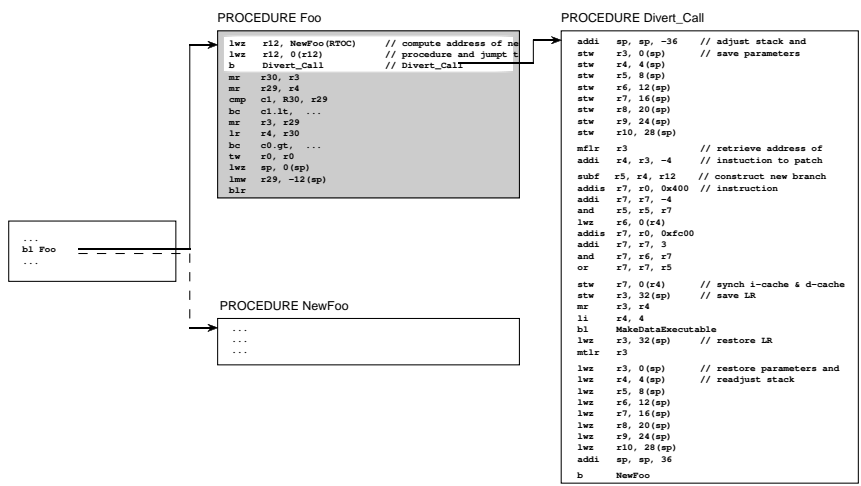}}
\caption{Diverting Procedure Calls}
\end{figure}

Unfortunately, this solution works only for direct branches and fails for indirect
calls via \emph{procedure variables}.  In the latter case, the entire code sequence preceding the call 
instruction needs to be analyzed to determine the memory location containing the procedure variable to be 
updated.  This may be simple in many cases, but may also be very complicated at times---especially when 
procedure variables are themselves passed as parameters to the calling procedure, or when the same procedure 
variable is stored in different locations at different times (which can happen when a procedure variable is 
spilled by the register allocator).  Furthermore, assignments to procedure variables can \emph{not} be 
corrected lazily because they do not involve a branch instruction.  Lastly, when lazy replacement is used, it 
is difficult to decide at which point the old code image can be discarded.  Unless a reference-counting 
mechanism is also implemented, it is unknown how many branches to the old image remain at any time.

Hence, a much better solution is to update dependencies \emph{instantaneously} as soon as the new code image 
has been generated.  This is the solution we have implemented.  In our architecture, code for a particular 
application is not allocated contiguously.  Rather, individual procedures are allocated separately as dynamic 
objects.  References to procedures, such as calls and procedure variables, are treated like 
object-pointers---that is, they are traced by the garbage collector.  Procedures that are no longer referenced 
by either a direct branch or a procedure variable can then be de-allocated fully automatically and memory 
fragmentation can be avoided.

In order to update dependencies, we extend the garbage collector by a \emph{translation mechanism} that accepts 
a list of translation tuples $(ptr_{old}, ptr_{new})$ as its input.  During the mark phase, the garbage 
collector automatically replaces all occurrences of $ptr_{old}$ by $ptr_{new}$.  Hence, in order to replace a 
procedure by a newly optimized replacement in our system, one constructs a translation tuple and calls the 
garbage collector.  Note that the garbage collector must trace the stack and all registers for this method to 
work correctly.  The advantage of this mechanism is that all references are updated at once, so that the full 
benefit of the optimization can be used immediately.  It also makes it possible to discard the old code image 
right away.  However, this technique is also not perfect.  It incurs the overhead of having to run the garbage 
collector, although this is not a serious problem in practice.

Besides the question of how to replace code images, we also have to address the issue
of \emph{when} to replace them.  This is particularly difficult in the case in which a procedure is replaced 
while it is active.  In this situation, replacing code in situ is virtually impossible because the original 
continuation point may no longer exist.  For example, a loop that had been partially executed before the 
replacement occurred could be unrolled in the new image.  Hence, replacement in situ is possible only at 
specific synchronization points that have been inserted artificially, but this limits optimization potential.

This particular problem goes away when the new code image is constructed in a different
location. Execution simply continues in the old version of the procedure, and the
new code isn't executed until the next invocation of the procedure. However, if the
procedure being replaced consists of a long-running task, the results of the applied
optimizations may never take effect at all.

There are also optimizations that require an immediate switch to the new code. For
example, we have implemented an optimization component that improves data cache locality
of pointer-centric program code. To this effect, our optimization uses
profiling information to build a temporal relationship graph of data-member accesses.
It then computes the optimal internal layout of data objects and recompiles the affected
procedures to access data members using this layout. Obviously, this optimization
mandates that at the same time the code is exchanged, all the existing data objects
are simultaneously also transformed into the new format. In our solution, it is the
garbage collector that performs both of these tasks. For this reason, the garbage collector
in our system is in fact also extensible.

As a consequence of wanting to explore optimizations such as to the last mentioned,
our implementation permits the substitution of procedures only when it can be guaranteed
that no active thread is executing them. This still rules out the replacement of
long-running tasks, but considerably reduces implementation complexity. Since our
system is structured around a central ``event loop'', such long-running tasks don't occur
in practice.

\section{Open Questions} 
Although our architecture is remarkably extensible and allows adding and removing
components at will, a flexible architecture also increases the risk that anomalies
arise from unexpected interactions between independently provided components. As
an example, a prefetching optimization might be composed of two interdependent optimization
phases: one that inserts prefetching instructions into the intermediate representation
(GSA) and a second one that implements a prefetching-aware instruction scheduler.
If the prefetching-aware scheduler is subsequently replaced by another scheduler
that does not handle prefetching instructions correctly, the overall system performance
might deteriorate. A further problem already alluded to above is the general compiler
phase-ordering problem: depending on the order in which optimizations are executed,
code of different quality is generated \cite{CC95}.

Solving these problems is difficult as it involves knowing all the dependencies among
individual components. Certain misconfigurations between independently developed
components can be detected and eliminated by using configuration management tools.
These tools require that individual components be specified using a special specification
language and then verify the system's consistency whenever it is modified. Specification
languages can be used to capture both structural and semantic dependencies.  \emph{Structural dependencies} 
between components can be captured by simple rules such as ``optimization phase X needs to be executed after 
optimization phase Y'' and ``optimization phase X needs to be executed directly before optimization phase Y'' 
\cite{Mon98}.  Capturing the \emph{semantic dependencies} is much more difficult.  It requires the use of a 
formal notation to describe the semantic actions of components that allows deriving
relations between optimizations, such as whether two optimizations can be combined, whether an optimization 
enables another optimization, or whether an optimization disables another optimization.

At this point in time, it is still unclear whether the semantic actions can be described
accurately enough in a formal notation to be of practical use in configuration management
tools. We believe that configuration management is an important issue that needs
to be addressed in order for fully extensible systems to become reliable and wide-spread
in use, but contend that this is an issue that is orthogonal to questions of architecture.
The component interaction dilemma is akin to the possible interference between dynamic
link libraries in a modern operating system, which hasn't prevented DLL's from appearing
in such systems. Our architecture solves the genuine problem of maintaining multiple
versions of an optimizing compiler at the core of a run-time system without affecting
the run-time system in this process; the configuration management problem is secondary
as long as the number of extension providers remains small.

\section{Related Work} 
Pioneering research in dynamic runtime optimization was done by Hansen \cite{Han74} who first described a fully 
automated system for runtime code optimization.  His system was similar in structure to our system---it was 
composed of a loader, a profiler, and an optimizer---but used profiling data only to decide when to optimize 
and what to optimize, not how to optimize.  Also, his system interpreted code prior to optimization, since 
load time code generation was too memory and time consuming at the time.

Hansen's work was followed by several other projects that have investigated the benefits of runtime 
optimization: the Smalltalk \cite{DS84} and SELF \cite{Hol94} systems that focused on the benefits of dynamic 
optimization in an object-oriented environment; ``Morph'', a project developed at Harvard University 
\cite{ZWG+97}; and the system described by the authors of this paper \cite{Kis96,KF96}.  Other projects have 
experimented with optimization at link time rather than at runtime \cite{Wal92:Systems}.  At link time, many 
of the problems described in this paper are non-existent.  Among them the decision when to optimize, what to 
optimize, and how to replace code.  However, there is also a price to pay, namely that it cannot be performed 
in the presence of dynamic loading.

Common to the above-mentioned work is that the main focus has always been on functional
aspects, that is how to profile and which optimizations to perform. Related to this
is research on how to boost application performance by combining profiling data and
code optimizations at compile time (not at runtime), including work on method dispatch optimizations for object-oriented
programming languages \cite{Ung86,HCU91}, profile-guided intermodular optimizations \cite{SW93,CMCH92}, code positioning techniques
\cite{PH90,CL96}, and profile-guided data cache locality optimizations \cite{CKJA98,CL98,KF98:RecordLayout}.

In contrast, the main emphasis of this paper is on architectural aspects of a system
service that provides dynamic runtime optimization. We believe that it is often the
architectural aspects that separate a curious research idea from a practical solution
that can be incorporated in a real-world system.

\section{Conclusion} 
We have presented an extensible and customizable architecture for dynamic runtime
optimization. Our system is flexible enough to accommodate a wide variety of current
and future optimization and profiling techniques. These can be provided by hardware
manufacturers as encapsulated ``plug-in'' components, permitting an incremental update
of the run-time system without involvement of the original software supplier.

\begin{acks} 
We would like to thank Peter Fr\"ohlich, Ziemowit Laski, and Christian Stork who provided many helpful comments.  
\end{acks}

\newpage

\bibliographystyle{Architecture}
\bibliography{Architecture}

\end{document}